\documentclass[a4paper, final]{article}

\usepackage[english]{babel}
\usepackage[T1]{fontenc}
\usepackage{amsmath}
\usepackage{amssymb}
\usepackage[dvips]{graphicx}
\usepackage{subfigure}
\usepackage{cases}
\usepackage{empheq}
\usepackage{bbm}
\usepackage{color}

\newtheorem{lemma}{Lemma}

\begin{document}

\title{Surface Waves in Almost Incompressible Elastic Materials}
\author{Kristoffer.\ Virta$^{1,*}$, Gunilla.\ Kreiss$^{1}$\\
\\
$^{1}$ Division of Scientific Computing,\\
Department of Information Technology, Uppsala.\\
$^{*}$Email: kristoffer.virta@it.uu.se}
\date{}
\maketitle

\begin{abstract}
A recent study shows that the classical theory concerning accuracy and points per wavelength is not valid for surface waves
in almost incompressible  elastic materials. The grid size must instead be proportional to $(\frac{\mu}{\lambda})^{(1/p)}$ to achieve a certain accuracy. Here $p$ is the order of accuracy the scheme and $\mu$ and $\lambda$ are the Lame parameters.  This accuracy requirement becomes very restrictive close to the incompressible limit  where $\frac{\mu}{\lambda} \ll 1$, especially for low order methods.
We present results concerning how to choose the number of grid points for 4th, 6th and 8th order summation-by-parts  finite difference schemes. The result is applied to Lambs problem in an almost incompressible material.   

\end{abstract}

\section{Introduction}
Consider the half - plane problem for the two - dimensional elastic wave equation in a homogeneous isotropic material. With time scaled to give unit density the displacement field $(u,v)$ is governed by 
\begin{equation}
	\begin{array}{ll}
		u_{tt} = \mu \Delta u + (\lambda + \mu)(u_x+v_y)_x,\\
		v_{tt} = \mu \Delta v + (\lambda + \mu)(u_x+v_y)_y,
	\end{array}
		(x,y) \in (-\infty,\infty) \times [0,\infty),t \geq 0,
	\label{eq:e1}
\end{equation}
where $\lambda > 0$ and $\mu > 0$ are the first and second Lame' parameters of the material. We assume that both Lame' parameters are constant. Initial data for $(u,v)$ and $(u_t,v_t)$ is given at $t = 0$. On the boundary $y = 0$ we consider conditions on the normal and tangential stresses
\begin{equation}
	\begin{array}{ll}
		v_y + \frac{\lambda}{\lambda + 2\mu} u_x = g_1(x,t),\\
		u_y + v_x = g_2(x,t),
	\end{array}
		y = 0, t > 0.
	\label{eq:e2}
\end{equation}
With $g_1 = g_2 = 0$ (\ref{eq:e2}) is called a traction free boundary condition. The elastic energy, a semi - norm of the solution to (\ref{eq:e1}), is given by
\begin{equation}
\label{eq:e3}
E(t) = \frac{1}{2} \int_{0}^{\infty} \int_{-\infty}^{\infty} \left(u_t^2 + v_t^2\right) + \lambda \left(u_x + v_y\right)^2 + \mu \left(2u_x^2+2v_y^2+\left(u_y+v_x\right)^2\right) dx dy.
\end{equation}
The elastic energy satisfies (see e.g., \cite{c1}, pp. 582 - 600)
\begin{equation}
\label{eq:e4}
\frac{d}{dt} E(t) = -\int_{-\infty}^{\infty} \left(v_t\left(\lambda + 2\mu\right)g_1 + u_t \mu g_2\right)_{y = 0} dx.
\end{equation}
In particular, with a traction free boundary condition the elastic energy is constant,
\begin{equation}
\label{eq:e5}
E(t) = E(0), t \geq 0, g_1  = g_2  = 0.
\end{equation}
It is well known that (\ref{eq:e1}) admits compressional and shear waves. This becomes transparent when considering the simpler set of equations equivalent to (\ref{eq:e1}), 
\begin{equation}
	\begin{array}{ll}
		\phi_{tt} = \left( \lambda + 2 \mu \right) \Delta \phi,\\
		H_{tt} = \mu \Delta H,
	\end{array}
	(x,y) \in (-\infty,\infty) \times [0,\infty),t \geq 0.
	\label{eq:e6}
\end{equation}
Here the equations for $\phi$ and $H$, governs the propagation of compressional and shear waves with phase velocities $\sqrt{\lambda + 2 \mu}$ and $\sqrt{\mu}$, respectively. The displacement field $(u,v)$ is obtained via
\begin{equation}
	\begin{array}{ll}
		u = \phi_x + H_y,\\
		v = \phi_y - H_x.
	\end{array}
	\label{eq:e7}
\end{equation} 
The boundary condition (\ref{eq:e2}) in terms of $\phi$ and $H$ becomes,
\begin{equation}
	\begin{array}{ll}
		\phi_{xx} + \phi_{yy} - \frac{2 \mu}{\lambda + 2 \mu} (\phi_{xx} + H_{xy}) = g_1(x,t),\\
		2 \phi_{xy} + H_{yy} - H_{xx} = g_2(x,t),
	\end{array}
	y = 0, t > 0.
	\label{eq:e8}
\end{equation}
For a discussion on how to arrive at (\ref{eq:e6}) - (\ref{eq:e8}) from (\ref{eq:e1}) - (\ref{eq:e2}) see \cite{c1}, pp. 273 - 278. The elastic wave equation with a traction free boundary condition also admits Rayleigh surface waves. These waves travel harmonically along the surface of the half - plane, whereas the amplitude decay exponentially into the domain. The phase velocity $c_R$ of the waves satisfies $c_R/\sqrt{\mu} < 1$. The exact value of the quotient depends on $\mu/\lambda$, but an approximation is given in \cite{c2} by $c_R/\sqrt{\mu} \approx (0.87 + 1.12 \nu)/(1+\nu) < 1$. Here $\nu = \lambda/2(\lambda + \mu)$. Hence, the Rayleigh surface waves always travel slower than both the compressional and shear waves. In many applications the period of the solution is given through boundary and internal forcing and can be considered as known. Then, as the phase velocity of a wave is defined by the ratio of its length and period, the shortest present wavelengths becomes proportional to $\sqrt{\mu}$. According to the classical theory in \cite{c3} an accurate numerical solution is obtained if the shortest wave length is not smaller than a constant number of grid sizes, where the constant depends on the order of accuracy of the numerical method. This predicts that the grid size should be proportional to $\sqrt{\mu}$. In a recent paper by H - O. Kreiss and N.A. Petersson \cite{c4} materials with $\mu \ll \lambda$ are studied. There it is shown that the classical theory is inadequate when simulating surface waves. Instead it is proved that the grid size must be proportional to $\left(\mu/\lambda\right)^{1/p}$ in order to achieve an accurate solution. Here $p$ is the order of accuracy of the numerical method. This requirement becomes very restrictive close to the incompressible limit $\mu / \lambda \ll 1$, especially for low order methods. The theory in \cite{c4} was supported by numerical experiments using 2nd and 4th order discretizations of (\ref{eq:e1}) - (\ref{eq:e2}) with $\mu / \lambda$ as small as $10^{-3}$. Another discretization of \eqref{eq:e1} - \eqref{eq:e2} was constructed in \cite{c5}. The discretization uses summation - by - parts (SBP) finite difference operators of orders $2,4,6,8$ \cite{c9,c10} to discretize the right hand side of \eqref{eq:e1}. The method uses the simultaneous - approximation - term (SAT) method \cite{c6} to approximate the boundary conditions \eqref{eq:e2}. By using the properties of the SBP operators stability of the resulting scheme was proven by constructing a discrete semi - norm of the discrete solution with the property of mimicking (\ref{eq:e4}). In particular, the discretization with a traction free boundary condition mimics \eqref{eq:e5} to machine precision. Accuracy and convergence of the discretization was verified by using a standing wave solution. In this paper we continue in the lines of \cite{c4} and use the code developed in \cite{c5} to further study simulation of surface waves in almost incompressible materials. In particular we study materials in which $\mu/\lambda < 10^{-3}$. 

\begin{figure}[htbp]
	\centering
	\includegraphics[scale=0.2]{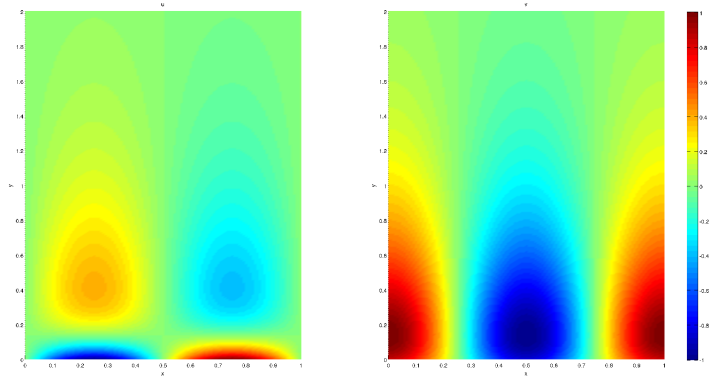}
	\caption{Plot of the $(u,v)$ components of the Rayleigh surface wave at $t=52.34$ for a material with $\lambda = 1$ and $\mu = 10^{-4}$. The $u-\mathrm{component}$ is shown to the right and the $v-\mathrm{component}$ to the left.}
	\label{fig:f1}
\end{figure}

\begin{figure}[htbp]
	\centering
	\includegraphics[scale=0.4]{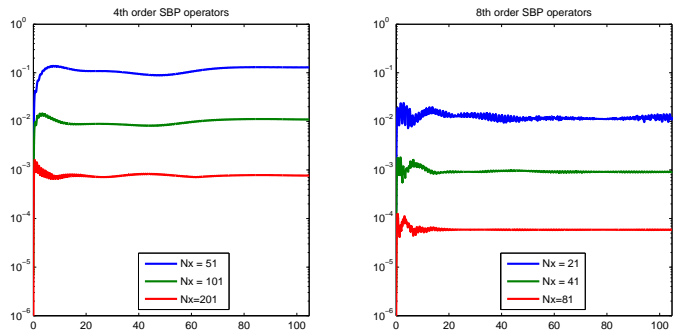}
	\caption{Relative max error as a function of time for the Rayleigh surface wave in a material with $\lambda = 1$ and $\mu = 10^{-4}$. Results from a scheme using 4th and 8th order SBP operators are shown on the left and right, respectively. The number of points per wavelength is increased from top to bottom. Note that the grids are finer for the computations using 4th order SBP operators.}
	\label{fig:f2}
\end{figure}

In the concluding section of \cite{c4} remarks are made on the use of methods of higher order than 4. It is there concluded that numerical experiments must be performed to evaluate how small $\mu/\lambda$ has to be to compensate for the higher complexity of higher order methods. As an introductory example we therefor let a Rayleigh surface wave propagate in the half - plane $y \geq 0$ with a traction free boundary condition at $y = 0$. The wave clings to the surface and decays exponentially in $y$, see Figure \ref{fig:f1}. We take $\lambda = 1$ and $\mu = 10^{-4}$. The solution is scaled such that the surface wave has unit wavelength. The resulting period of the solution is then $T = 104.678$. In the numerical experiment the $x$ - direction is made $1$ - periodic. The performance of methods using 4th and 8th order SBP operators are then compared. We use $N_x$ points per surface wave length and compute until time $T/2$. In Figure \ref{fig:f2} the relative max error as a function of time is displayed for the different methods on a series of finer grids. Note that to achieve a relative max error of at most $5\%$ the method using 4th order SBP operators require $101$ grid points per surface wavelength. This is approximately 10 times the number of points predicted by the classical theory. The figure showing the results for the method using 8th order SBP operators shows that only 21 grid points per surface wavelength is needed to make the relative max error less than $5\%$. 

The originality of this work follows in Section \ref{sec:s3} and \ref{sec:s4}. Section \ref{sec:s3} presents numerical tests on the performance of higher order methods with $\mu/\lambda$ as small as $10^{-6}$. These results are used in section \ref{sec:s4} to estimate the number of points per smallest wavelength needed to accurately approximate a version of Lambs problem (\cite{c7}) in a almost incompressible material. In an appendix we derive an analytic expression for the Rayleigh surface wave and discuss its sensitivity to a boundary truncation error in a numerical approximation. This presentation is analogous to the one given in \cite{c4} but differs in that the theory is obtained via the equations \eqref{eq:e6} - \eqref{eq:e8} rather than \eqref{eq:e1} - \eqref{eq:e2}. Concluding remarks are given in section \ref{sec:s5}.          
\section{The numerical method}
\label{sec:s2}
The elastic wave equation on the second order form \eqref{eq:e1} was discretized in \cite{c5}. To approximate spatial operators high order SBP operators were used. In \cite{c5} it was shown how to impose a traction free boundary condition weakly with the SAT technique. A Dirichlet condition was imposed strongly by injecting data at the boundary. Stability of the numerical scheme was proved with the energy method by showing that the discrete system satisfies a discrete energy estimate mimicking \eqref{eq:e5}. The discretization and proof of the energy estimate was done for general SBP operators without any restrictions on the order of accuracy. In this paper we consider numerical schemes constructed with $2p$ - th order SBP operators \cite{c9,c10} for $p = 2,3,4$. Although termed $2p$ - th order accurate the local order of accuracy is only $p$ at a constant number of points in the vicinity of the boundary of the domain. It has been shown in \cite{c13} for a discretization of the Schr\"odinger equation using $2p$ - th order SBP operators the global order of accuracy is $p + 2$. This was also observed in the numerical experiments of \cite{c5}.    

In the numerical experiments we impose periodic boundary conditions at $x = \pm L_x$ in the $x$ - direction. At a distance $L_y$ below the traction free surface at $y = 0$ we either impose a Dirichlet condition, when the exact solution is known, or use the perfectly matched layer (PML) constructed in \cite{c8} to absorb outgoing waves. 
\begin{figure}
	\centering
	\includegraphics[scale=0.5]{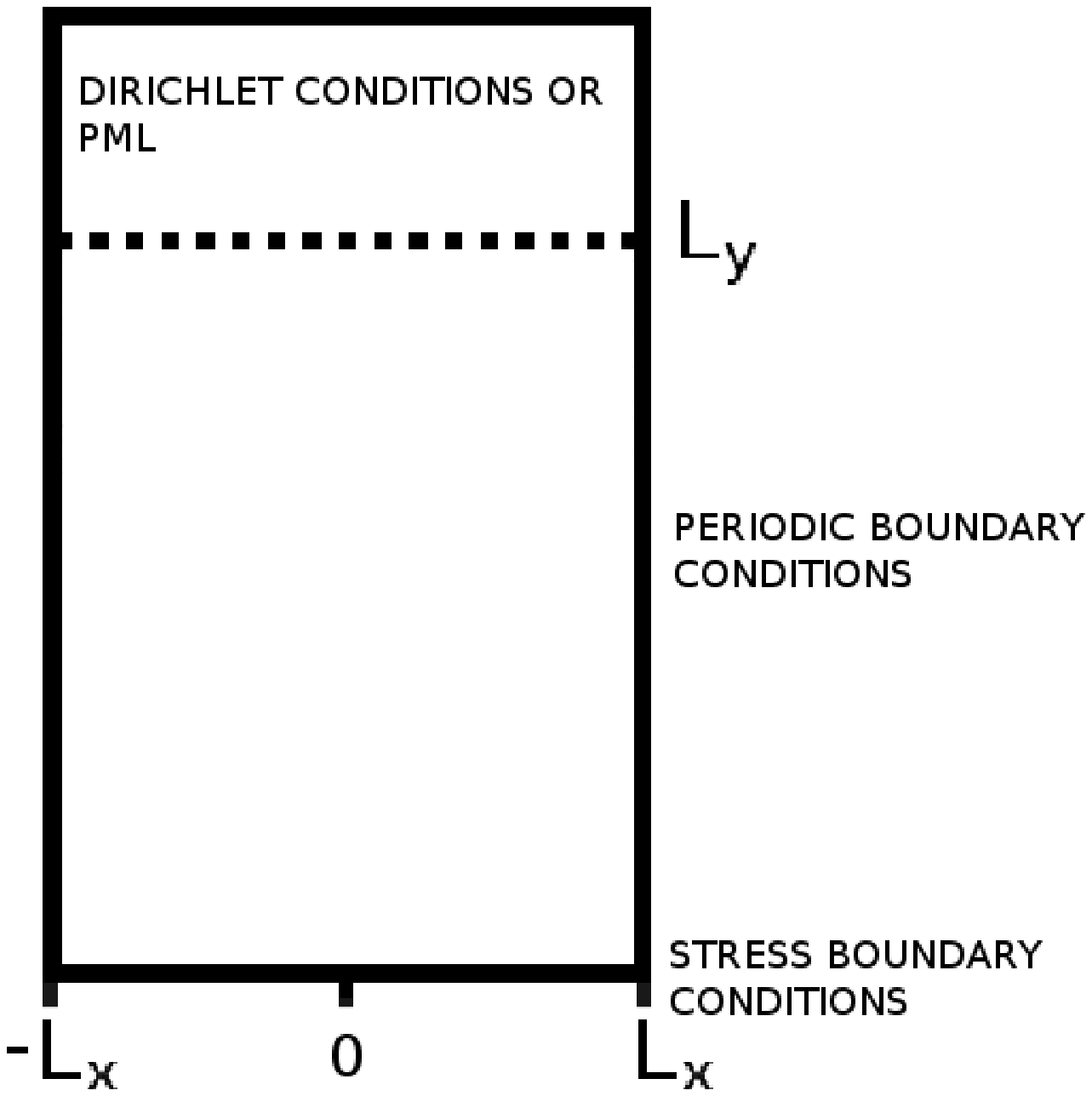}
\end{figure}

The discretization of spatial derivatives with SBP operators, enforcing of a traction free boundary condition with the SAT technique and a Dirichlet condition at $y = L_y$ results in a semi - discrete system of the type 
\begin{equation}
	\label{eq:s2e1}
	\begin{bmatrix}
		\mathbf{u} \\
		\mathbf{v}
	\end{bmatrix}_{tt} = \frac{1}{h^2}\mathbf{Q} \begin{bmatrix}
		\mathbf{u} \\
		\mathbf{v}
	\end{bmatrix}.
\end{equation}
Here $\mathbf{u}$ and $\mathbf{v}$ are vectors with approximative values of $u$ and $v$ at the grid points of the computational domain and $h$ is the grid size. $\mathbf{Q}$ is a matrix with elements independent of $h$.
To discretize \eqref{eq:s2e1} in time we use the 4th order time stepping scheme of \cite{c12}. This scheme was designed for a system of the type \eqref{eq:s2e1} in that it is not rewritten to a system first order in time. In \cite{c12} it was shown that if a time step $k$ is chosen as 
\begin{equation}
	\label{eq:s2e2}
	k = h \frac{C}{\left(\|\mathbf{Q}\|_\infty \|\mathbf{Q}\|_1\right)^{1/4}} 
\end{equation}
stability is guaranteed provided that a discrete energy estimate exists. Here $C$ is a constant depending on the order of accuracy of the spatial discretization. Using the PML results in a system of the type 
\begin{equation}
	\label{eq:s2e3}
	\begin{bmatrix}
		\mathbf{u}_{tt} \\
		\mathbf{v}_{tt} \\
		\mathbf{w}_{t}
	\end{bmatrix} = \mathbf{P} \begin{bmatrix}
		\mathbf{u} \\
		\mathbf{v} \\
		\mathbf{w}
	\end{bmatrix}.
\end{equation}
Here $\mathbf{w}$ contains auxiliary variables arising from the addition of the PML.  In the presence of a PML the system \eqref{eq:s2e3} is rewritten as a first order system in time and the classical Runge - Kutta 4 scheme is used to integrate in time.  
\section{A numerical study for different values of $\mu/\lambda$}
\label{sec:s3}
To perform reliable numerical computations it is of importance to know the number of grid points per surface wavelength needed to obtain a certain accuracy in an approximate solution. In this section we study surface waves in materials with varying $\mu/\lambda$. We are interested in the performance of schemes using SBP operators of different orders to discretize (\ref{eq:e1}) - (\ref{eq:e2}). The numerical study is performed as follows. For a given value of $\mu/\lambda$ we determine the number of grid points, $P_R$, per surface wavelength needed to achieve a relative max error of at most $5\%$ after having propagated for 10 periods in time. As the number of grid points per surface wavelength needed for accuracy is proportional to $\left(\lambda / \mu\right)^{1/p}$ where $p$ is the order of accuracy of the method, high order of accuracy is expected to become more influential as $\mu/\lambda$ decreases. The higher the order of the scheme the more computational effort is required. For this reason execution times are recorded. It is then possible to conclude how small $\mu/\lambda$ must be to compensate for the higher complexity of higher order schemes. 

We use (\ref{eq:e18}) and (\ref{eq:e7}) to derive an analytic expression of a Rayleigh surface wave. We are computing in real arithmetic, therefor we use the real part of the displacement field $(u,v)$,   
\begin{equation}
	\label{eq:e39}
	Re\left(\begin{pmatrix} u & v \end{pmatrix}^T\right)
	= 
	A_2 \begin{pmatrix}
		\left(-\xi e^{-\alpha y} + \frac{\beta^2+\xi^2}{2 \xi} e^{-\beta y}\right) \sin(\xi (x-c_R t))\\
    		\left(-\alpha e^{-\alpha y} + \frac{\beta^2+\xi^2}{2 \beta} e^{-\beta y}\right) \cos(\xi (x - c_R t))
	\end{pmatrix}, A_2 \in \mathbb{R}. 
\end{equation}
Here the phase velocity $c_R$ is given by (\ref{eq:e19}). We keep the wavelength fixed at $L_R = 1$ by choosing $\xi = 2 \pi$. The values of $\alpha$ and $\beta$ then follows from (\ref{eq:e10}). The constant $A_2$ is arbitrary but we take $A_2 = -\frac{1}{2 \pi}$. For simplicity we keep $\lambda = 1$ fixed. The period, $T$, of the solution is then proportional to $1/\sqrt{\mu}$. We consider a domain which is periodic in the $x$ - direction. The computational domain is chosen to contain exactly one wavelength of the solution. At the boundary $y = 0$ a traction free boundary condition is imposed. The time step is chosen according to \eqref{eq:s2e2}. The computational domain is truncated at $y = 10$ by imposing exact boundary data given by the exact solution (\ref{eq:e39}). The numerical computations are made on a single Intel Xenon W3680 3.33 Ghz processor. Results for different values of $\mu/\lambda$ are reported in Table 2. We see that for $10^{-3} \leq \mu/\lambda \leq 10^{-2}$ the schemes using 6th and 8th order SBP operators perform similarly whereas the scheme using 4th order operators need significantly more computational time to achieve a $5\%$ relative error, in particular for the case $\mu = 10^{-3}$. The errors obtained with the 4th order method are of the same magnitude as those obtained in \cite{c4} for the same values of $\mu/\lambda$. For $10^{-5} \leq \mu/\lambda \leq 10^{-4}$ the 6th order scheme is clearly disadvantageous compared to the 8th order scheme, it uses more than 4 times the amount of time to get a relative error of maximum $5\%$ for $\mu/\lambda$ in this interval. For $\mu / \lambda \leq 10^{-6}$ the required number of points per surface wavelength used by the 8th order scheme has increased very much above the value predicted by the classical theory and for such materials even higher order methods would be needed for an efficient numerical method. These computations verifies the theory of \cite{c4} for schemes of higher accuracy than 4 and predicts how small $\mu / \lambda$ must be for the different higher order methods to be more efficient when surface waves are present in simulations.              

\begin{table}[htbp]
	\begin{center}
		\begin{tabular}{|l|l|l|l|l|l|l|l|}
		\hline
		Case & $P_R$ &$e_4$ & $T_4$ & $e_6$ & $T_6$ & $e_8$ & $T_8$ \\
		\hline
		$\mu = 10^{-2}$ & 13 & $1.4 \times 10^{0}$  & $1$  & $1.3 \times 10^{-1}$ & $4$ & $\color{red} 1.9 \times 10^{-2}$ & $\color{red}7$ \\
		$T = 10.474$    & 25 & $1.5 \times 10^{-1}$ & $11$ & $\color{red} 9.0 \times 10^{-3}$ & $\color{red}31$ & - & - \\
		                & 49 & $\color{red} 1.1 \times 10^{-2}$ & \color{red}$80$ & - & - & - & - \\
		\hline
		$\mu = 10^{-3}$ & 13 & $5.7 \times 10^{0}$  & $4$ & $6.6 \times 10^{-1}$ & $11$ & $ 7.5 \times 10^{-2}$ & $19$ \\
		$T = 33.104$    & 25 & $9.4 \times 10^{-1}$ & $37$ & $\color{red} 2.7 \times 10^{-2}$ & $\color{red}96$ & $\color{red} 4.3 \times 10^{-3}$ & $\color{red}173$ \\
			        & 49 & $8.8 \times 10^{-2}$ & $251$ & - & -  & - & - \\
			   	& 97 & $\color{red} 6.3 \times 10^{-3}$ & $\color{red}1990$ & - & - & - & - \\
		\hline
		$\mu = 10^{-4}$ & 13 &$4.8 \times 10^{0}$   & 17 & $3.5 \times 10^{0}$ & 36 & $ 3.0 \times 10^{-1}$ & $71$ \\
		$T = 104.678$   & 25 & $5.0 \times 10^0$    & 144 & $1.4 \times 10^{-1}$  & 318  & $\color{red} 6.6 \times 10^{-3}$ & $\color{red} 545$ \\
				& 49 & $6.3 \times 10^{-1}$ & 1078 & $\color{red} 5.2 \times 10^{-3}$ & $\color{red}2936$ & - & -\\
				& 97 & $5.3 \times 10^{-2}$ & 6407  & -  & - & - & - \\
				&193 & $\color{red} 3.7 \times 10^{-3}$ & \color{red} 79000 & - & - & - & -\\
		\hline
		$\mu = 10^{-5}$ & 13 & $5.1 \times 10^{0}$  & 36 & $4.2 \times 10^0$ & 114  & $2.4 \times 10^{-0}$  & 196 \\
		$T = 331.020$   & 25 & $5.4 \times 10^{0}$  & 452 & $8.8 \times 10^{-1}$ & 990  & $\color{red} 1.8 \times 10^{-2}$  & \color{red} 1746 \\
				& 49 & $3.4 \times 10^{0}$  & 3460 & $\color{red} 2.4 \times 10^{-2}$ & \color{red} 7397 & - & -  \\
				& 97 & $0.4 \times 10^{-1}$ &  32241 & - & - & - & -  \\
				&193 & $\color{red} 3.5 \times 10^{-2}$ &  $\color{red}270870$ & - & - & - & - \\
		\hline
		$\mu = 10^{-6}$ &  13 &-& - & $ 5.2 \times 10^0$ & 513 & $ 3.7 \times 10^{0}$  & 1043 \\
		$T = 1046.778$  &  25 &-& - & $ 4.5 \times 10^{0}$ &4288 & $ 1.3 \times 10^{-1}$& 9993 \\
				&  49 &-& - & $ 1.8 \times 10^{-1}$ &  39233 & $\color{red} 1.9 \times 10^{-3}$ & \color{red} 78775   \\
				&  97 &-& - & $\color{red} 4.4 \times 10^{-3}$  & \color{red} 322546  & -  & - \\
		\hline
		\end{tabular}
		\label{tab:t2}
		\caption{The leftmost column displays the value of $\mu/\lambda$ and the period, $T$, of the solution. The rest of the columns report the number of grid points per surface wavelength and the corresponding relative max errors, $e_p$ and execution times, $T_p$ for schemes using $p$-th order SBP operators.}
	\end{center}
\end{table}
\section{Application: Lambs problem in almost incompressible material}
\label{sec:s4}
\begin{figure}[htbp]
	\centering
	\subfigure[Numerical solution]{
  		\includegraphics[width=0.42\linewidth]{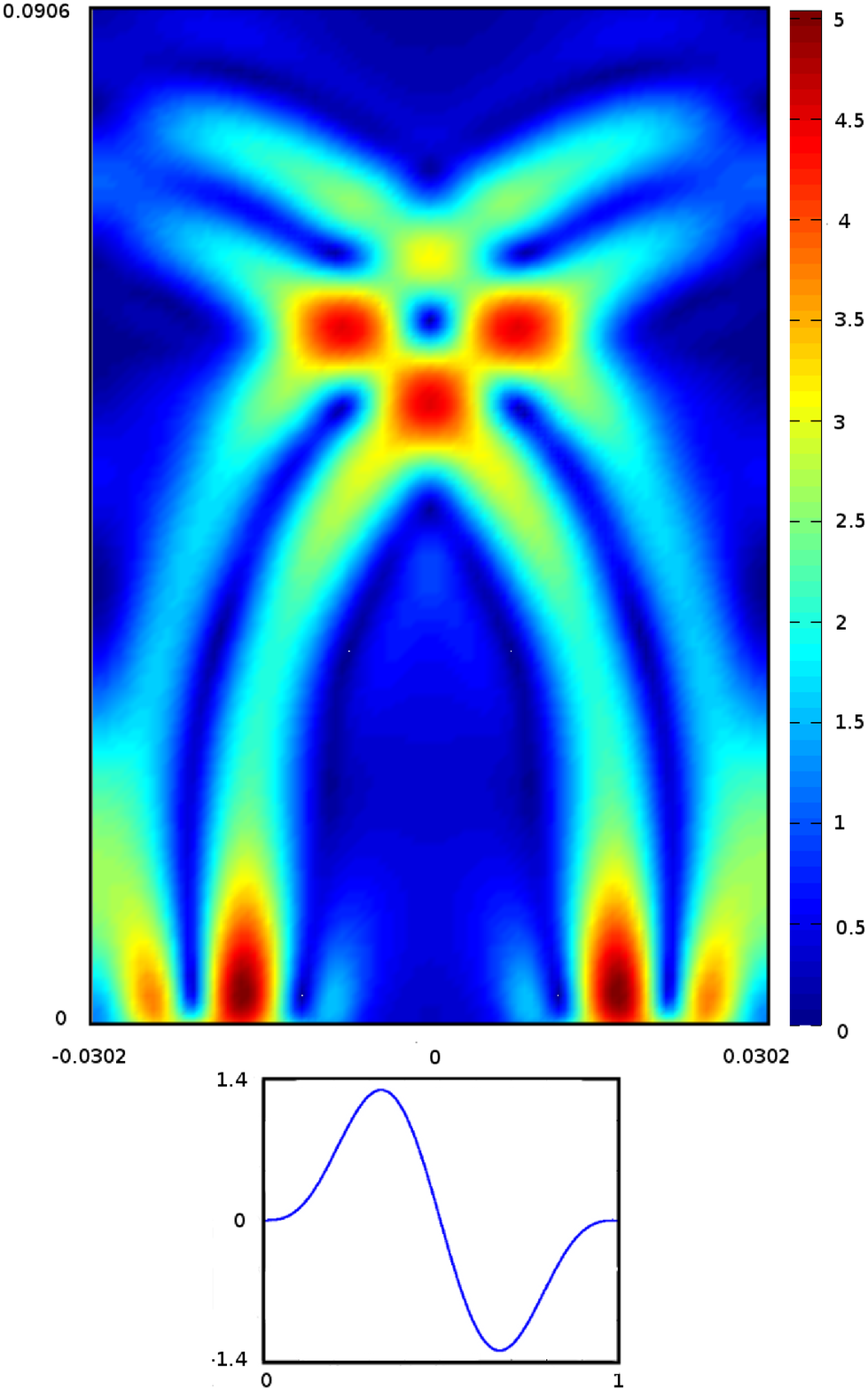}
 	\label{fig:f3}
	}
	\subfigure[Relative error]{
  		\includegraphics[width=0.42\linewidth]{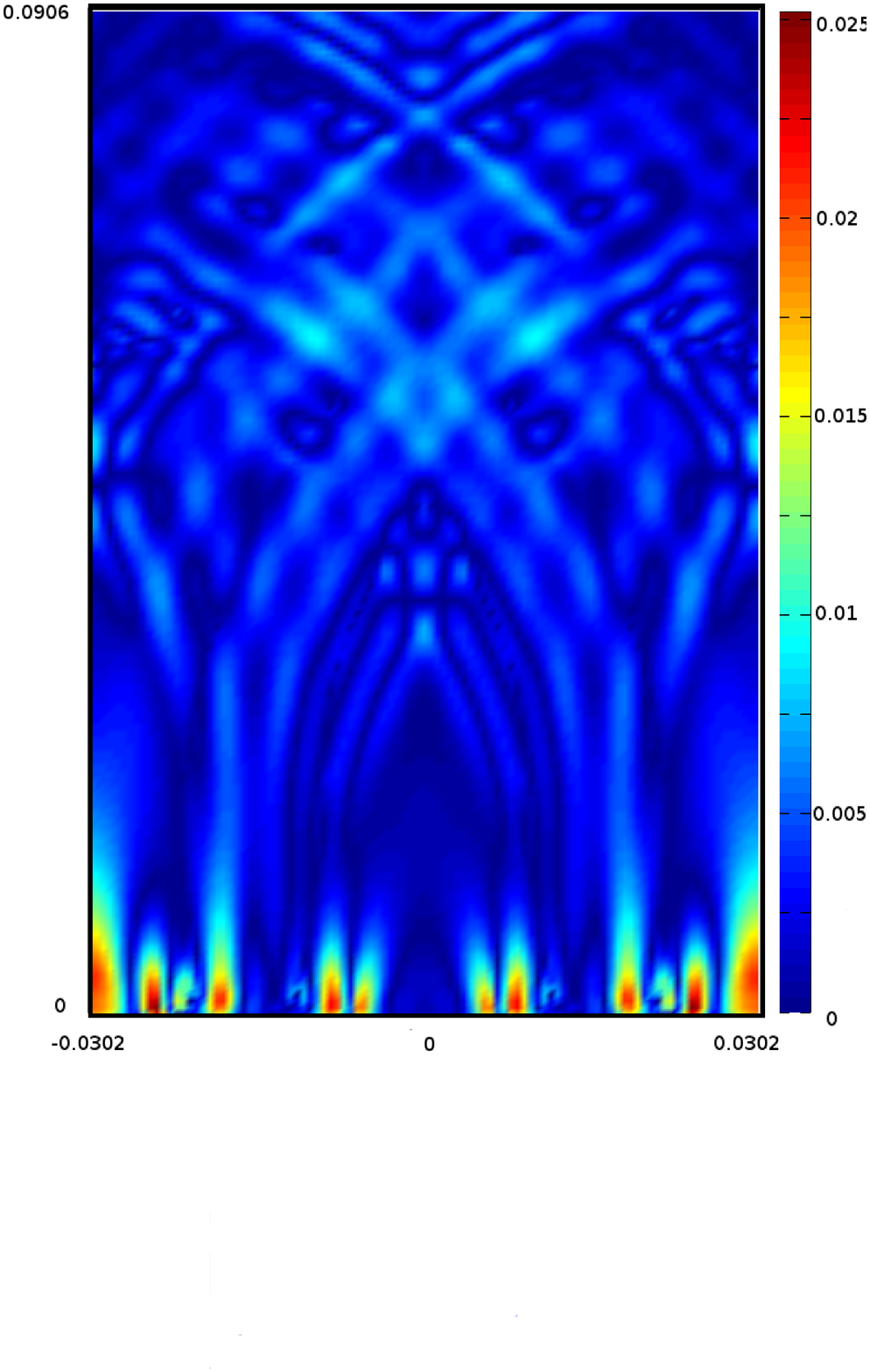}
 	\label{fig:f4}
	}
	\caption{The numerical solution (a) and the relative error (b) at time $t=3.2$}
\end{figure}

We solve a version of Lambs problem \cite{c7} in which the surface of a half - space is subjected to a periodic array of line sources with loading normal to the surface. Lamb proved in \cite{c7} that under these conditions compressional, shear and Rayleigh waves are generated. The stress forcing of (\ref{eq:e2}) is
\begin{equation*}
	\label{eq:e40}
	\begin{array}{ll}
		g_1(x,t) = f(t)\delta(x-k M), M > 0, k = 0, \pm 1, \dots,\\
		g_2(x,t) = 0,
	\end{array}
\end{equation*} 
where $M$ is the distance between the sources and $\delta$ the Dirac delta function. We let $f$ be the wavelet given by
\begin{equation*}
	\label{eq:e41}
	f(t) = \left\{\begin{array}{l}
			\sin(2 \pi \omega t) - \frac{1}{2} \sin(4 \pi \omega t),~ 0 \leq t \leq \frac{1}{\omega}\\
		        0,~ \mathrm{else}.
		  \end{array}\right.
\end{equation*}  
$f$ is shown as an inset in Fig \ref{fig:f3} with $\omega = 1$.
With $\lambda = 1, \mu = 10^{-3}$ the Rayleigh phase velocity becomes $c_R \sqrt{\mu} = 0.0302$. The highest significant frequency with $\omega = 1$ in the time function $f$ is $2$. The corresponding shortest wavelength of the Rayleigh wave is $L_{\mathrm{min}} = \frac{c_R\sqrt{\mu}}{2} = 0.0151$. We choose the domain $[-2L_{\mathrm{min}},2 L_{\mathrm{min}}]\times[0,6 L_{\mathrm{min}}]$, $M = 4L_{\mathrm{min}}$ and solve numerically until time $t = 3.2$. Figure \ref{fig:f3} shows the magnitude of the displacement field. Periodic boundary conditions are applied at the vertical boundaries and the domain is truncated above with a perfectly matched layer \cite{c8}. To estimate the required number of points per wavelength to achieve a relative max error of at most $5\%$ with a 6th order method we consult Table 2 to conclude that 25 points per wavelength should suffice. To ascertain this claim a reference solution with 200 points per shortest wavelength is constructed. As a comparison a solution using 10 points per shortest wavelength, a quantity predicted by the classical theory to yield a relative error lower than $5\%$, is also computed. The results presented in Table \ref{tab:t3} verifies the claim for this application. Figure (\ref{fig:f4}) shows the relative error in the magnitude of the displacement field at time $t = 3.2$. It is interesting to see that the main bulk of the error is seen to be located in the vicinity of the surface. This is in accordance with the theory presented in \cite{c4}, which predicts that the Rayleigh waves are much more sensitive to discretization errors than the shear and pressure waves.     
\begin{table}[htbp]
	\begin{center}
            	\begin{tabular}{lll}
            		$P$ & $e_6$  \\
			\hline
            		10 & $7.5 \times 10^{-1}$ \\
			\hline
            		25 & $2.7 \times 10^{-2}$  \\
            	\end{tabular}
		\caption{Points per shortest wavelength and corresponding relative max error.}
            	\label{tab:t3}	
            \end{center}      
\end{table}   
\section{Conclusions}
\label{sec:s5}
We have studied numerical difficulties in the simulation of surface waves in almost incompressible elastic materials. The work as been greatly influenced by the theory of H-O. Kreiss and N.A. Petersson in \cite{c4}. Here they showed that the number of grid points per wavelength of the surface wave needed for accuracy is proportional to $\left(\mu/\lambda\right)^{1/p}$, where $p$ is the order of accuracy of the method. This is opposing the classical theory which suggest a proportionality to $\sqrt{\mu}$. This requirement becomes more restrictive as the elastic material becomes more incompressible, $\mu/\lambda \rightarrow 0$. In this work we have used a SBP + SAT discretization of the elastic wave equation in a half - plane to study surface waves in materials in which $\mu/\lambda$ is as small as $10^{-6}$. The main goal was to investigate how small the quotient $\mu/\lambda$ must be to compensate for the higher complexity of higher order methods. In particular we have used methods of orders higher than 4. The results of this study was then used in an application where we numerically solved a version of Lambs problem in an almost incompressible material.

A synopsis of the existence of a Rayleigh surface wave and its sensitivity to boundary truncation was given in an appendix. This presentation is analogous to the one given in \cite{c4} with the difference that the results where developed from the formulation \eqref{eq:e6} - \eqref{eq:e8} rather than the formulation \eqref{eq:e1} - \eqref{eq:e2}. In the case of elastic wave propagation in two half - planes in welded contact the existence of Stoneley interface waves with much similarity to the Rayleigh surface wave can be proved \cite{c11}. In a ongoing study the authors aim to investigate numerical difficulties in the simulation of the Stoneley interface wave. It is then believed that an approach similar to the one given in the appendix may be fruitful. 
\appendix
\section{The Rayleigh surface wave and sensitivity to boundary truncation errors}
\subsection{The Rayleigh surface wave}
Consider the half - plane problem \eqref{eq:e6} - \eqref{eq:e8} with a traction free boundary condition, $g_1 = g_2 = 0$. We examine the existence of solutions of the type
\begin{equation}
	\label{eq:e9}
	\begin{array}{ll}
		\phi = f(y)e^{st+i\xi x}, & H = h(y)e^{st+i\xi x},\\
		|f|_\infty < \infty, & |h|_\infty < \infty, 
	\end{array}
	\xi \in \mathbb{R} \setminus \{0\}.
\end{equation}
Inserting (\ref{eq:e9}) into (\ref{eq:e6}) we get,
\begin{equation}
	\label{eq:e10}
	\begin{array}{ll}
		f^{\prime \prime} - \alpha^2 f = 0, \alpha^2 = \left(\xi^2+\frac{s^2}{\lambda + 2 \mu}\right),\\
		h^{\prime \prime} - \beta^2 h = 0, \beta^2 = \left(\xi^2+\frac{s^2}{\mu}\right).
	\end{array}
\end{equation}
The solution to (\ref{eq:e10}) is,
\begin{equation}
	\label{eq:e11}
	\begin{array}{ll}
		f = A_1 e^{\alpha y} + A_2 e^{-\alpha y},\\
		h = B_1 e^{\beta y} + B_2 e^{-\beta y}. 
	\end{array}
\end{equation}
$\phi$ and $H$ of (\ref{eq:e9}) then becomes,
\begin{equation}
	\label{eq:e12}
	\begin{array}{ll}
		\phi = A_1 e^{i\xi x + \alpha y + st} + A_2 e^{i\xi x - \alpha y + st},\\
		H = B_1 e^{i\xi x + \beta y + st} + B_2 e^{i\xi x - \beta y + st}.
	\end{array}
\end{equation}
Letting $A_1 = B_1 = 0$ and inserting the expressions (\ref{eq:e12}) for $\phi$ and $H$ into the boundary conditions (\ref{eq:e8}) we get,
\begin{equation}
	\label{eq:e13}
	\begin{array}{ll}
		\left(\beta^2 + \xi^2\right)A_2 + 2 i \beta \xi B_2 = 0\\
		-2 i \alpha \xi A_2 + (\beta^2 + \xi^2) B_2 = 0
	\end{array}
\end{equation}
The linear system (\ref{eq:e13}) has a solution if and only if its determinant is zero,
\begin{equation}
	\label{eq:e14}
	\begin{array}{ll}
		D = \left(\beta^2 + \xi^2\right)^2 - 4 \alpha \beta \xi^2 = 0.
	\end{array}
\end{equation}
Using the expressions (\ref{eq:e10}) for $\alpha$ and $\beta$ we can write (\ref{eq:e14}) in the form
\begin{equation}
	\label{eq:e15}
	\begin{array}{ll}
		D = -4\xi^4 \left(\sqrt{\frac{s^2}{\left(\lambda + 2 \mu\right)\xi^2} + 1} \sqrt{\frac{s^2}{\mu\xi^2} + 1} - \left(\frac{s^2}{2 \xi^2 \mu} + 1\right)^2 \right) = -4\xi^4 \varphi(\tilde{s}) = 0,
	\end{array}
\end{equation}
where 
\begin{equation}
	\label{eq:e16}
	\varphi(\tilde{s}) = \sqrt{1+\tilde{s}^2}\sqrt{1+\frac{\mu \tilde{s}^2}{\lambda + 2 \mu}} - \left(1+\frac{\tilde{s}^2}{2}\right)^2, \tilde{s} = \frac{s}{|\xi|\sqrt{\mu}}.
\end{equation}
Since we exclude $\xi = 0$, the zeros of the determinant (\ref{eq:e15}) are the solutions of $\varphi(\tilde{s}) = 0$. The function $\varphi(\tilde{s})$ was investigated in \cite{c4} its properties can be summarized in the following lemma, 
\begin{lemma}
\label{lemma:l1}
The function $\varphi(\tilde{s})$ has exactly three roots $\tilde{s} = 0$ and $\tilde{s} = \tilde{s}_0 = \pm i\omega_0$. $\omega_0$ depends on $\mu/\lambda$ and $0 < \omega_0 < 1$. Furthermore, $|\varphi(\tilde{s}_0)^\prime|$ is bounded away from zero for all $\mu/\lambda$.   
\end{lemma}
Values of $\tilde{s}_0$ and $|\varphi(\tilde{s}_0)^\prime|$ for some values of $\mu/\lambda$ are calculated in Table 1.
\begin{table}[htbp]
	\begin{center}
		\begin{tabular}{|l|l|l|}
			\hline
			$\mu/\lambda$ & $\tilde{s}_0$ & $|\varphi(\tilde{s}_0)^\prime|$\\
			\hline
			$1$ & $0.9194 i$ & $1.0616$\\
			\hline
			$10^{-2}$ & $0.9547 i$ & $2.1576$\\
			\hline
			$10^{-4}$ & $0.9553 i$ & $2.1927$\\
			\hline
			$0$ & $0.9553 i$ & $2.1930$\\
			\hline
		\end{tabular}
		\label{table:tab1}
		\caption{The roots $\tilde{s}_0$ of $\varphi(\tilde{s})$ and magnitude of $\varphi(\tilde{s}_0)^\prime$ for some values of $\mu/\lambda$.}
	\end{center}
\end{table}
Using the fact $0 < \omega_0 < 1$ in the expression (\ref{eq:e16}) for $s$ we get
\begin{equation}
	\label{eq:e17}
	\begin{array}{ll}
		\alpha^2 = \left(\xi^2 - \frac{\xi^2\mu \omega_0^2}{\lambda + 2 \mu}\right) > \xi^2\left(1 - \frac{\mu}{\lambda + 2 \mu}\right) > 0,\\
	        \beta^2 = \left(\xi^2 - \frac{\xi^2\mu \omega_0^2}{\mu}\right) > \xi^2\left(1 - \frac{\mu}{\mu}\right) = 0.  
	\end{array}
\end{equation}
Hence, 
\begin{equation}
	\label{eq:e18}
	\begin{array}{ll}
		\phi = A_2 e^{-\alpha y } e^{i\xi \left(x \pm \omega_0 \sqrt{\mu} t\right)},\\
	        H = B_2 e^{-\beta y } e^{i\xi \left(x \pm \omega_0 \sqrt{\mu} t\right)},
	\end{array}
	 \alpha, \beta > 0, \xi \in \mathbb{R} \setminus \{0\} , \omega \in \mathbb{R}
\end{equation}
represents Rayleigh surface waves with amplitude that decays exponentially in the $y$ - direction. The waves travel harmonically along the $x$ - axis with phase velocity 
\begin{equation}
	\label{eq:e19}
	c_R = \omega_0 \sqrt{\mu}.
\end{equation}
Note that $A_1$ and $B_1$ necessarily vanishes, otherwise the solutions would have an unbounded amplitude for increasing $y$. In case of the root $\tilde{s} = 0$ (\ref{eq:e13}) gives 
$A_2/B_2 = 1/i$ so that 
\begin{equation}
	\notag
	\begin{array}{ll}
		\phi = A_2 e^{i\xi \left(x - y\right)},\\
		H = i A_2 e^{i\xi \left(x - y\right)}.
	\end{array}
\end{equation}
The relation (\ref{eq:e7}) then gives $u = v = 0$. That is, the root $\tilde{s} = 0$ corresponds to a displacement field that vanishes everywhere. 
\subsection{Sensitivity to boundary truncation errors}
The truncation errors arising from a discretization of the traction free boundary condition can be thought of as a perturbation of the homogeneous boundary condition by introducing non-zero boundary forcing functions $g_1,g_2$ in (\ref{eq:e2}) and (\ref{eq:e7}). Typically $g_1$ and $g_2$ depends on derivatives of the continuous solution and the grid size. We again consider a solution of the form (\ref{eq:e12}) with $A_1 = B_1 = 0$, 
\begin{equation}
	\label{eq:e20}
	\begin{array}{ll}
		\phi = A_2 e^{i\xi x - \alpha y + st},\\
		H = B_2 e^{i\xi x - \beta y + st}.
	\end{array}
\end{equation}
Inserting this solution into the now inhomogeneous boundary conditions (\ref{eq:e8}) we get after some algebra,   
\begin{eqnarray}
	\label{eq:e21}
		\left(1+\frac{\tilde{s}^2}{2}\right) A_2 + i \sqrt{1+\tilde{s}^2} B_2 = \frac{\left(\lambda + 2 \mu\right)g_1}{2 \mu \xi^2}\\
	\label{eq:e22}
		i \sqrt{1+\frac{\mu \tilde{s}^2}{\lambda + 2 \mu}} A_2 - \left(1 + \frac{\tilde{s}^2}{2}\right) B_2 = -\frac{g_2}{2\xi^2},
\end{eqnarray} 
where $\tilde{s} = \frac{s}{|\xi|\sqrt{\mu}}$. The determinant of this system is $\varphi(\tilde{s})$ defined by (\ref{eq:e16}). Hence, (\ref{eq:e21}) - (\ref{eq:e22})  becomes singular exactly at the roots of $\varphi(\tilde{s})$. Eliminating $B_2$ from (\ref{eq:e22}) and inserting it into (\ref{eq:e21}) gives,
\begin{equation}
	\label{eq:e23}
	\varphi(\tilde{s}) A_2 = -\frac{\lambda + 2 \mu}{2 \xi^2} \left(\frac{g_1}{\mu}\left(1+\frac{\tilde{s}^2}{2}\right) - i\sqrt{1+\tilde{s}}g_2\right).
\end{equation}
Let the grid size be $h$. Discretizing (\ref{eq:e2}) with a second order accurate method the principal part of the truncation errors becomes 
\begin{equation}
	\label{eq:e24}
	\begin{array}{ll}
		g_1 = \tau_{11} h^2 u_{xxx} + \tau_{12} h^2 v_{yyy},\\
		g_2 = \tau_{21} h^2 v_{xxx} + \tau_{22} h^2 u_{yyy},
	\end{array} 
	y = 0.
\end{equation}
By (\ref{eq:e7}),
\begin{equation}
	\label{eq:e25}
	\begin{array}{ll}
		g_1 = \tau_{11} h^2 \left(\phi_{xxxx} + H_{yxxx}\right) + \tau_{12} h^2 \left(\phi_{yyyy}-H_{xyyy}\right),\\
	        g_2 = \tau_{21} h^2 \left(\phi_{yxxx}-H_{xxxx}\right) + \tau_{22} h^2 \left(\phi_{xyyy} + H_{yyyy}\right),
	\end{array}
	y = 0. 
\end{equation}
Using (\ref{eq:e20}) the boundary forcing functions becomes,
\begin{equation}
	\label{eq:e26}
	\begin{array}{ll}
		g_1 = \tau_{11} h^2 \left(\xi^4 A_2 - i \xi^3 \beta B_2\right) + \tau_{12} h^2 \left(\alpha^4 A_2 - i\xi \beta^3 B_2\right),\\
	        g_2 = \tau_{21} h^2 \left(-i \xi^3 \alpha A_2 - \xi^4 B_2 \right) + \tau_{22} h^2 \left(i \xi \alpha^3 + \beta^4 B_2\right).
	\end{array}
\end{equation}
Since the right-hand side of (\ref{eq:e21}) is proportional to $g_1/\mu$ while the right - hand side of (\ref{eq:e22}) is independent of $\mu$, for $\mu \ll \lambda$ the main effect comes form $g_1$. To simplify we therefor assume that $g_2 = 0$. Using (\ref{eq:e21}) with $g_2 = 0$ we eliminate $b_2$ from $g_1$,
\begin{equation}
\label{eq:e27}
	\begin{array}{ll}
		g_1 = \tau_{11} h^2 \left(\xi^4 - i \xi^3 \beta \frac{i\left(1+\frac{\mu \tilde{s}^2}{\lambda + 2\mu}\right)}{1+\frac{\tilde{s}^2}{2}}\right)A_2 + \tau_{12} h^2 \left(\alpha^4 - i\xi \beta^3 \frac{i\left(1+\frac{\mu \tilde{s}^2}{\lambda + 2\mu}\right)}{1+\frac{\tilde{s}^2}{2}}\right) A_2\\
	g_2 = 0.
	\end{array}
\end{equation}
The solution formula (\ref{eq:e23}) can then be written in the form of an eigenvalue problem
\begin{equation}
	\label{eq:e28}
	\varphi(\tilde{s}) A_2 = \theta(\tilde{s}) A_2,
\end{equation}
where
\begin{equation}
	\label{eq:e29}
	\theta(\tilde{s}) = -\frac{\left(\lambda + 2 \mu\right)h^2\xi^2}{2 \mu} \left( \left( \tau_{11} \frac{\beta}{|\xi|} + \tau_{12} \frac{\beta^3}{|\xi|^3} \right) \left(1+\frac{\mu\tilde{s}^2}{\lambda+2\mu}\right) + \left(\tau_{11} + \tau_{12} \frac{\alpha^4}{\xi^4}\right) \left(1+\frac{\tilde{s}^2}{2}\right) \right). 
\end{equation}
This eigenvalue problem arises as a consequence of introducing truncation errors in a discretization of the traction free boundary condition. The eigenvalues $\tilde{s}$ of this problem determines the phase velocities of surface waves in a numerical solution of (\ref{eq:e1}) - (\ref{eq:e2}). The phase velocity of the Rayleigh surface wave was determined by the roots $\tilde{s}_0$ of the function $\varphi(\tilde{s})$. We now investigate how sensitive the difference $\tilde{s}-\tilde{s}_0$ is to truncation errors. We have for $\mu/\lambda \ll 1$,
\begin{equation}
	\label{eq:e30}
	\tilde{s}_0^2 \approx - 0.9, \varphi(\tilde{s}_0)^\prime \approx \pm 2.12 i, \frac{\alpha}{|\xi|} = \sqrt{1+\frac{\mu\tilde{s}^2}{\lambda + 2 \mu}} \approx 1, \frac{\beta}{|\xi|} = \sqrt{1 + \tilde{s}^2}  \approx 0.3.  
\end{equation}
Therefor 
\begin{equation}
	\label{eq:e31}
	\theta(\tilde{s}_0) \approx -\frac{\lambda h^2 \xi^2}{2 \mu}\left(0.3 \tau_{11} + 0.027 \tau_{12} + 0.55\left(\tau_{11} + \tau_{12}\right)\right).
\end{equation}
Taylor expanding (\ref{eq:e28}) about $\tilde{s}_0$ gives
\begin{equation}
	\label{eq:e32}
	\left(\tilde{s} - \tilde{s}_0\right) \varphi^\prime(\tilde{s}_0) \approx \theta(\tilde{s}_0).
\end{equation}
We get, 
\begin{equation}
	\label{eq:e33}
	\tilde{s} - \tilde{s}_0 \approx \frac{\theta(\tilde{s}_0)}{\varphi(\tilde{s}_0)} \approx \mp i\frac{\lambda h^2 \xi^2}{2 \mu}\frac{\left(0.3 \tau_{11} + 0.027 \tau_{12} + 0.55\left(\tau_{11} + \tau_{12}\right)\right)}{2.12}.
\end{equation}
To achieve a relative error in the phase velocity of size $\epsilon$, with $0 < \epsilon \ll 1$ we must choose the grid size $h$ such that
\begin{equation}
	\label{eq:e34}
	\frac{\lambda h^2 \xi^2|\tau|}{\mu} = \epsilon, \tau = i \frac{\left(0.3 \tau_{11} + 0.027 \tau_{12} + 0.55\left(\tau_{11} + \tau_{12}\right)\right)}{4.24}.
\end{equation}
If the computational grid has $P_R$ points per surface wavelength $L_R = 2\pi/|\xi|$ we get
\begin{equation}
	\label{eq:e35}
	h = \frac{L_R}{P_R}, h |\xi| = \frac{2\pi}{P_R}, P_R = 2 \pi \sqrt{\frac{|\tau|}{\epsilon}\frac{\lambda}{\mu}}.
\end{equation}
That is, as $\mu/\lambda \rightarrow 0$ the number of points per surface wavelength must be proportional to $\sqrt{\lambda/\mu}$ to maintain an relative error in the phase velocity of $\epsilon$. For a $p$-th order method the leading order truncation errors terms are 
\begin{equation}
	\label{eq:e36}
	g_1 = \tau_{11}^\prime h^p \frac{\partial^{p+1}u}{\partial x^{p+1}} + \tau_{12}^\prime h^p \frac{\partial^{p+1}v}{\partial y^{p+1}}
\end{equation}
and (\ref{eq:e34}) is replaced by 
\begin{equation}
	\label{eq:e37}
	\frac{\lambda h^p \xi^2|\tau^\prime|}{\mu} = \epsilon.
\end{equation}
The number of grid points required to maintain an error in the phase velocity of $\epsilon$ now becomes
\begin{equation}
	\label{eq:e38}
	P_R = 2 \pi \left(\frac{|\tau^\prime|}{\epsilon}\frac{\lambda}{\mu}\right)^{1/p}.
\end{equation}
Hence, as $\mu/\lambda \rightarrow 0$ the number of grid points per surface wave length grows much slower for larger $p$. 

\end{document}